\documentclass[12pt]{iopart}
\usepackage{graphicx}
\usepackage{iopams}

\def\PRL{Phys. Rev. Lett.\ }

\begin{document}

\title[PHENIX Highlights]{PHENIX Highlights.}

\author{Anthony D. Frawley\dag~for the PHENIX\footnote[3]{For the full PHENIX Collaboration author list 
and acknowledgments , see Appendix ``Collaborations'' of this volume.} collaboration}
\address{\dag Department of Physics, 
Florida State University,
Tallahassee, FL 32306}

\ead{frawley@fsuhip.physics.fsu.edu}

\begin{abstract}

Recent highlights of measurements by the PHENIX experiment at RHIC are presented.
 
\end{abstract}





The purpose of this talk was to highlight some of the most interesting recent results from PHENIX.
These include comparisons of d+Au and Au+Au data for identified hadrons at midrapidity, Au+Au two-particle 
jet correlations with and without particle identification, studies of nonrandom $p_T$ fluctuations and
HBT radii in Au+Au, the first measurement of direct photons in Au+Au collisions at RHIC, measurements 
of hadrons in the muon arms in d+Au,
$J/\psi$ measurements in d+Au reactions, and $\phi \rightarrow ee$ and $\phi \rightarrow KK$ comparisons
in d+Au reactions. All of the results are described in more detail in other contributions to these proceedings. 

\begin{figure}[htb]
 \centering
  \includegraphics[width=0.52\textwidth]{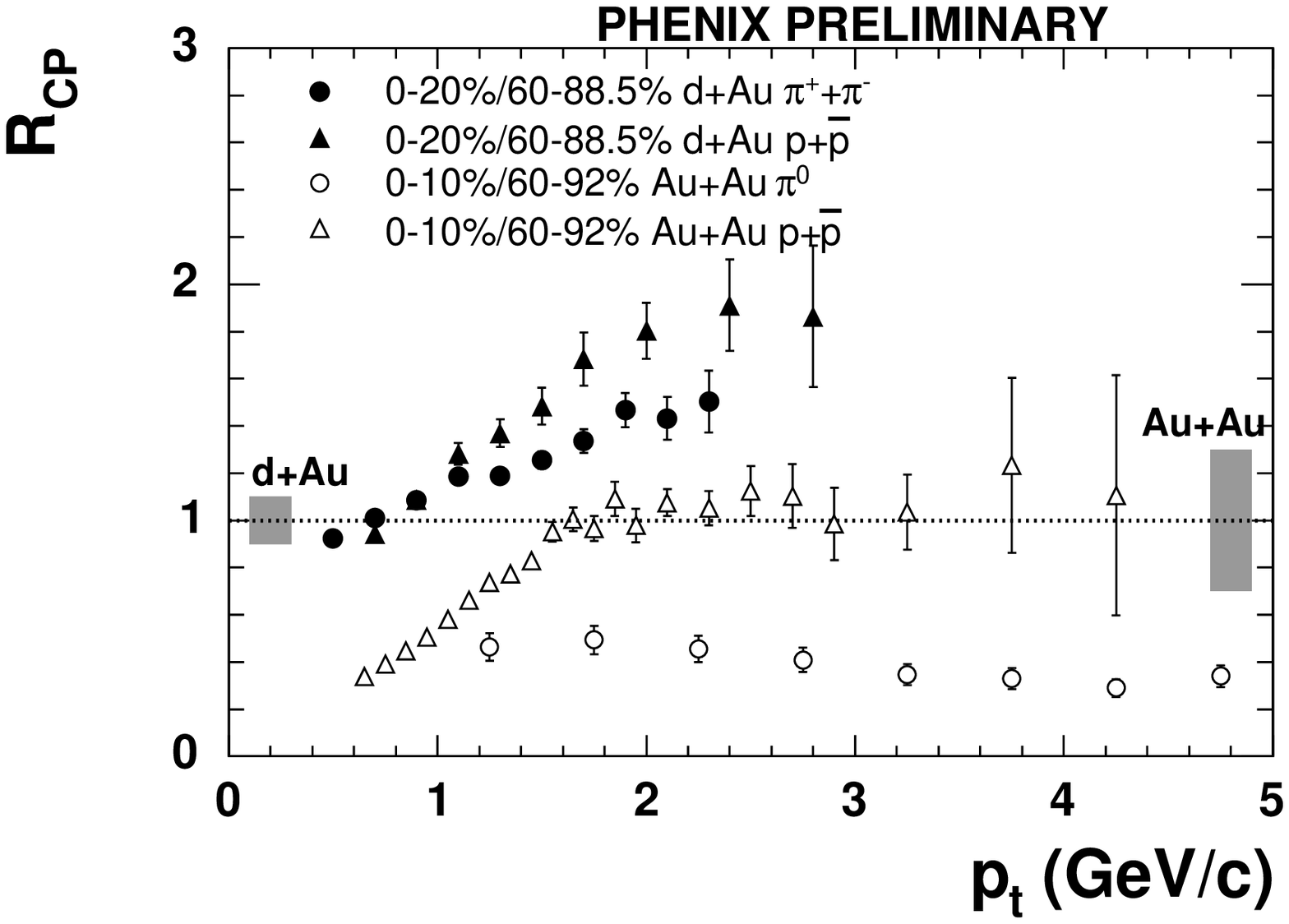}
  \includegraphics[width=0.47\textwidth]{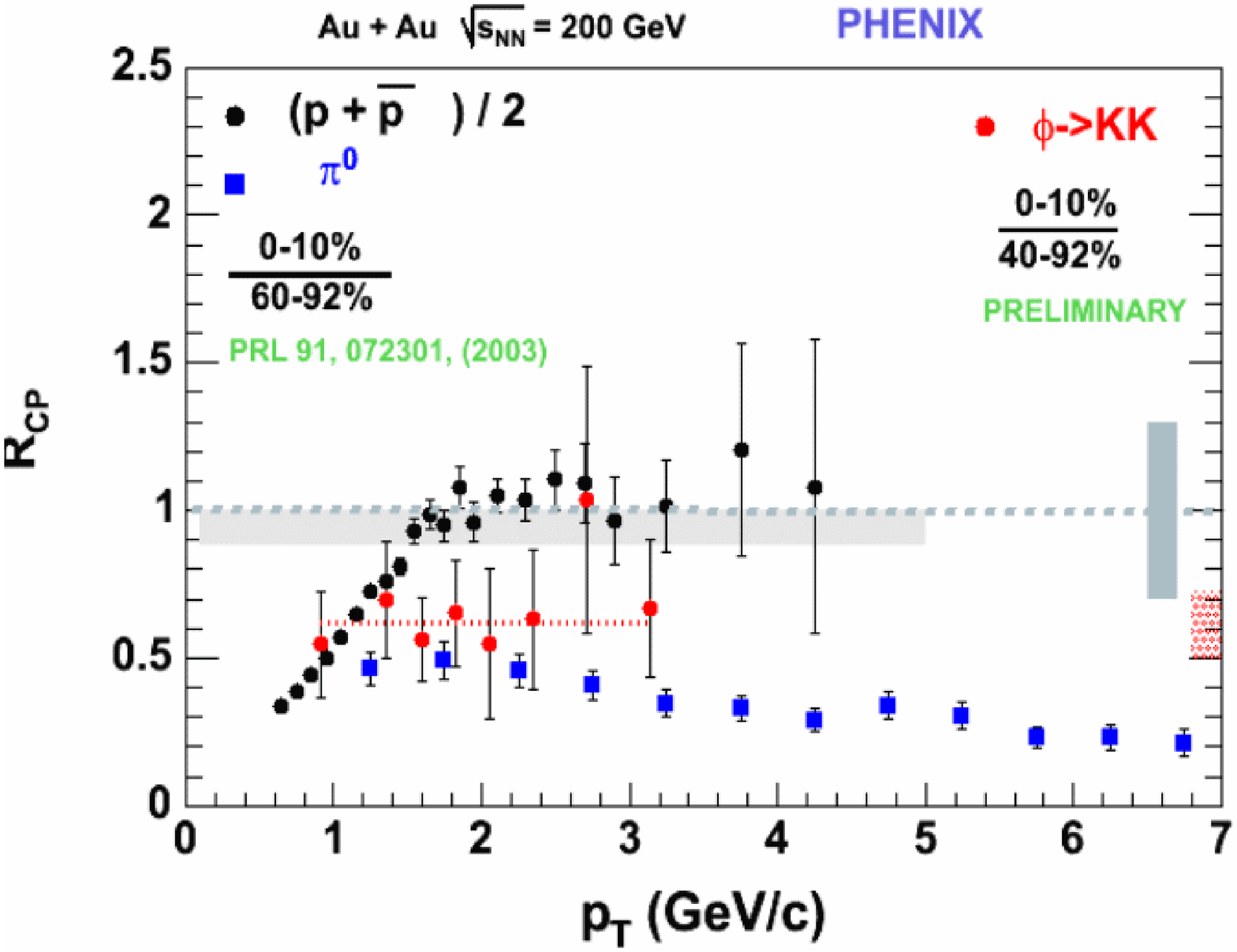}
  \caption{Left: pions are three times more heavily suppressed than 
protons in central Au+Au collisions at 200 GeV, while pion yields are only 25\% lower than proton yields 
in d+Au in d+Au collisions. Right: $\phi$ meson suppression is similar to $\pi^0$ suppression in central 
Au+Au collisions.}
  \label{fig:identified_particles}
\end{figure}

A comparison versus centrality of unidentified charged hadron data and $\pi^0$ data for Au+Au and 
d+Au collisions shows that hadron yields are enhanced at high collision centrality in d+Au, but 
suppressed as centrality increases in Au+Au~\cite{phenix_highpt}. Thus entrance channel effects can
not explain the strong hadron suppression in central Au+Au collisions. One very interesting feature is
that the nuclear modification factor for unidentified charged hadrons is significantly larger than 
that for pions for $p_T$ of about 2-5 GeV/c. As can be seen in figure~\ref{fig:identified_particles}, 
identified charged particle data for Au+Au 
collisions~\cite{phenix_pid} show that protons are unsuppressed in this momentum range.

To investigate whether the observed enhancement of protons over pions in the medium $p_T$ range in 
central Au+Au collisions could be due to a difference in Cronin effect, identified charged particle 
data for p+p, d+Au and Au+Au collisions have been compared~\cite{phenix_pid}, as shown in 
figure~\ref{fig:identified_particles}. 
Although the Cronin enhancement in d+Au is 20\% larger
for protons, the effect is too small to explain the factor of three difference in yields in 
central Au+Au collisions. Thus the enhancement must be attributed to the conditions formed in central
Au+Au collisions. In the same figure, it is shown that the 
$\phi$ suppression is comparable with that of the pion, providing evidence that the difference in 
suppression between protons and pions is related to quark number, not mass~\cite{phi_rcp}.

\begin{figure}[htb]
 \centering
  \includegraphics[width=0.45\textwidth]{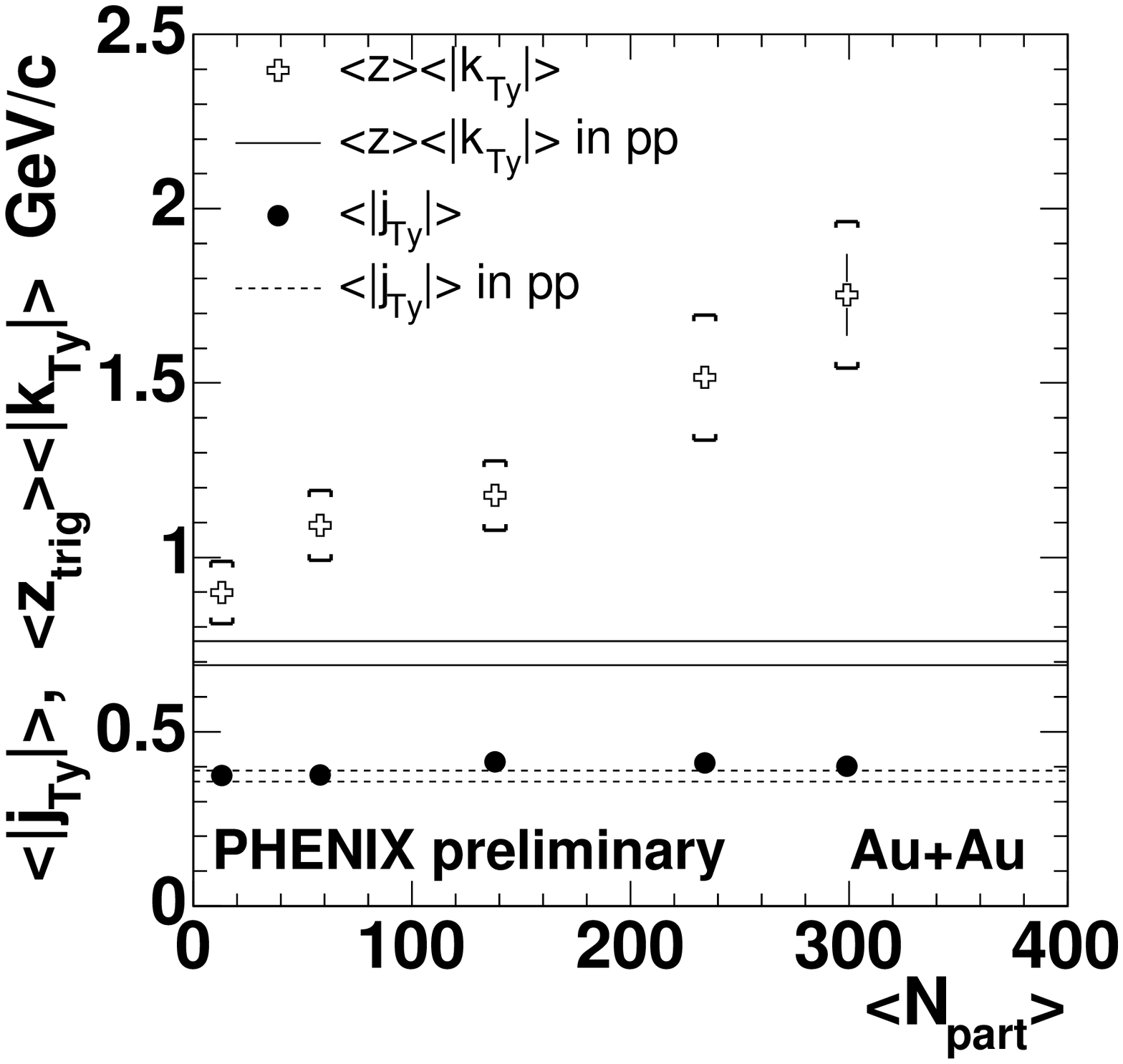}
  \includegraphics[width=0.45\textwidth]{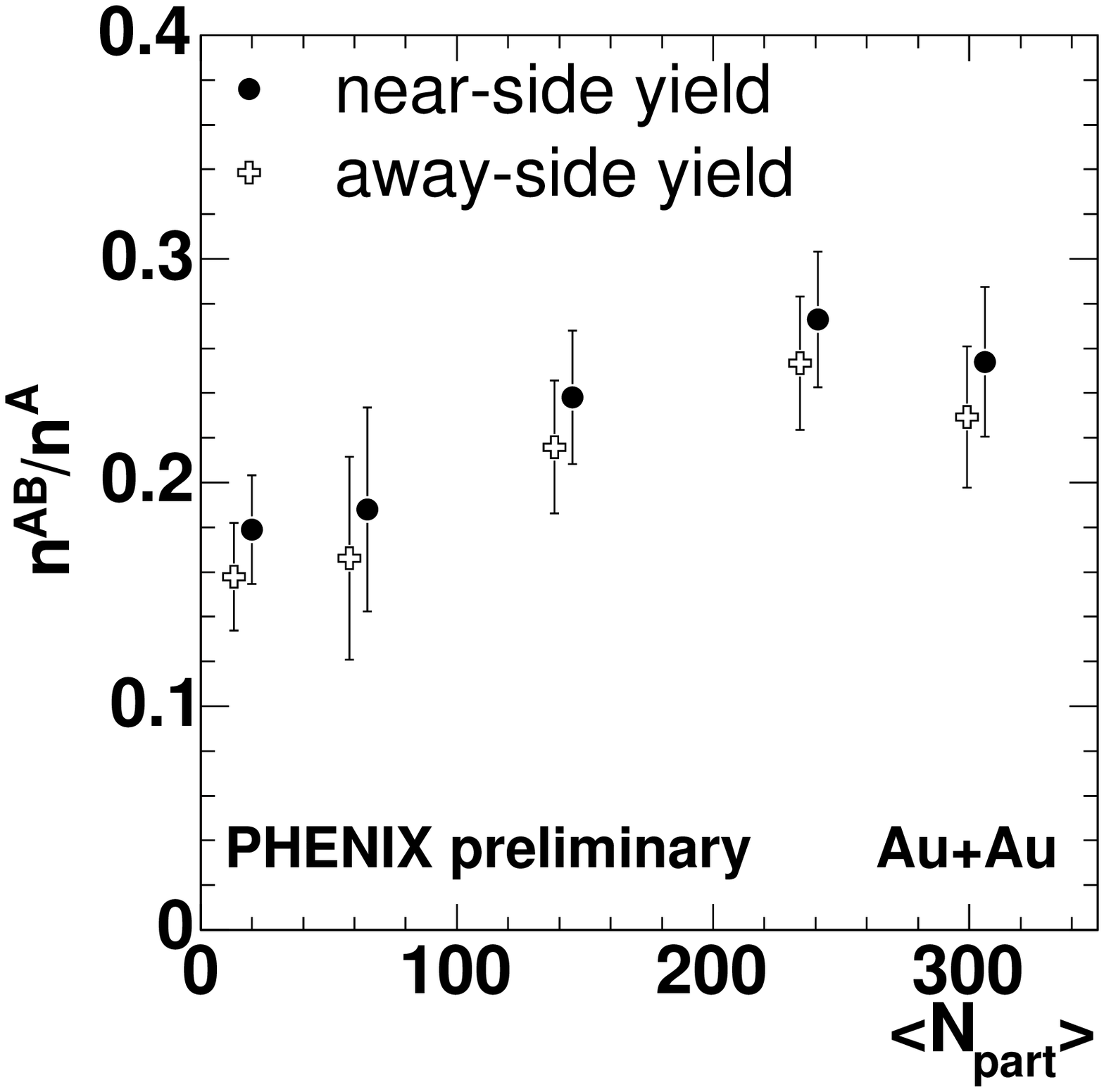}
  \caption{Left: Widths of the near side and away side peaks in the two-particle azimuthal 
correlations as a function of centrality. The lines show the values observed in p+p collisions. Right: 
The near and away side associated particle yields as a function of centrality. The trigger particles have
$2.5 < p_T < 4.0$ GeV/c and the associated particles have $1.0 < p_T < 2.5$ GeV/c.}
  \label{fig:jan_jets}
\end{figure}

PHENIX has studied the properties of jets in p+p, d+Au and Au+Au collisions at 200 GeV using two-particle
azimuthal correlations in which the trigger particle is assumed to be the leading particle from a high 
$p_T$ jet~\cite{jets} and the second particle is either from the same jet or the recoil jet. 
Figure~\ref{fig:jan_jets} shows the 200 GeV Au+Au centrality dependence of the measured widths 
(characterized by $<j_{Ty}>$) of the near side jet correlation function and 
(characterized by $<z><K_{Ty}>$) the far side jet correlation function. The near side width is seen 
to be essentially constant while the far side width increases markedly with centrality. Also shown
is the centrality dependence of the near and away side associated yields per trigger particle,
integrated over the full (gaussian) distribution in each case. The associated yields are seen to be
very similar to each other and to rise somewhat with centrality. There is no evidence for the disappearance 
of the away-side jet in this $p_T$ range, contrary to previous observations in correlation measurements
between trigger and associated particles with somewhat larger $p_T$~\cite{STAR_awaysidejet}.

\begin{figure}
\center
\includegraphics[width=0.48\textwidth]{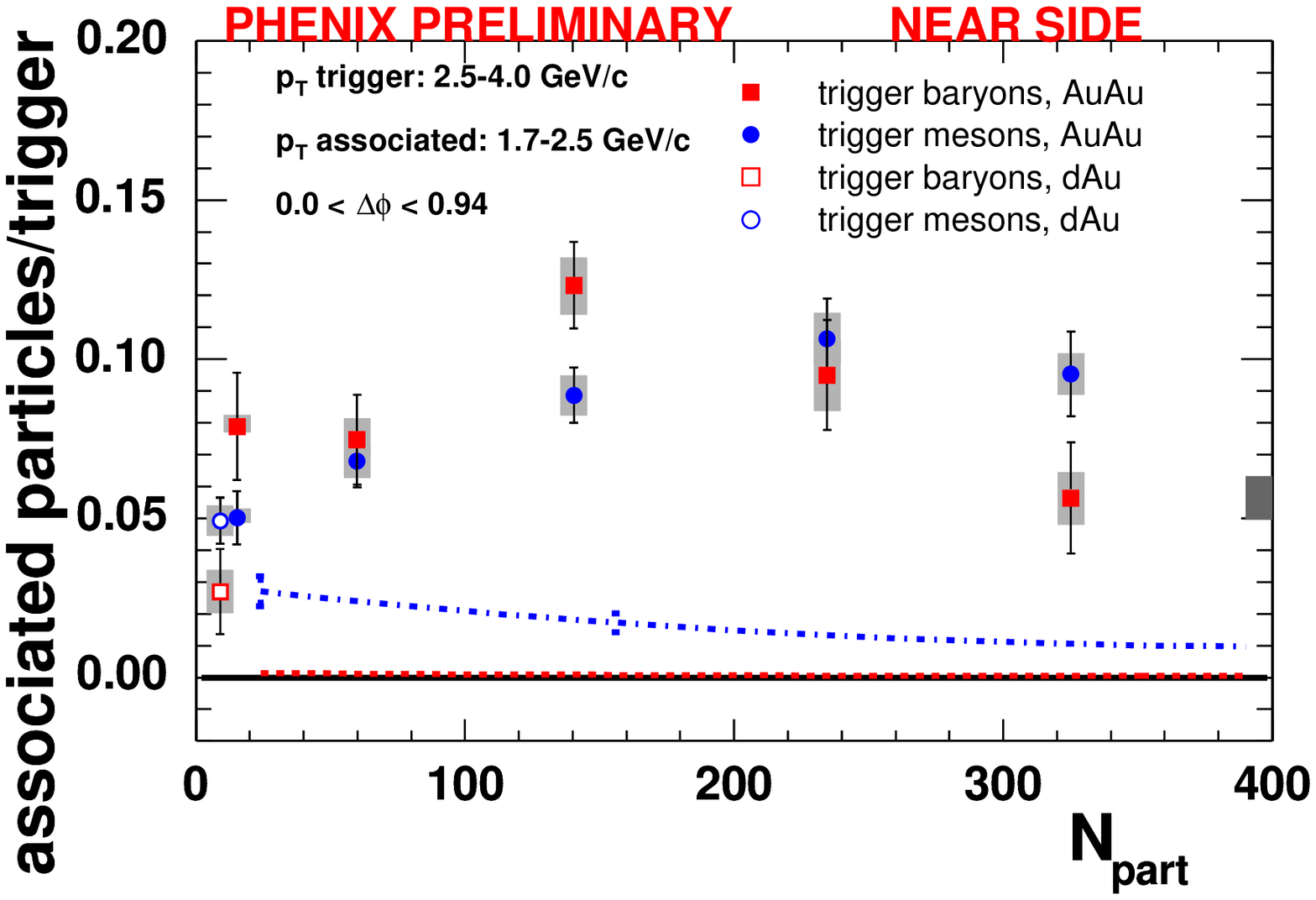}
\includegraphics[width=0.48\textwidth]{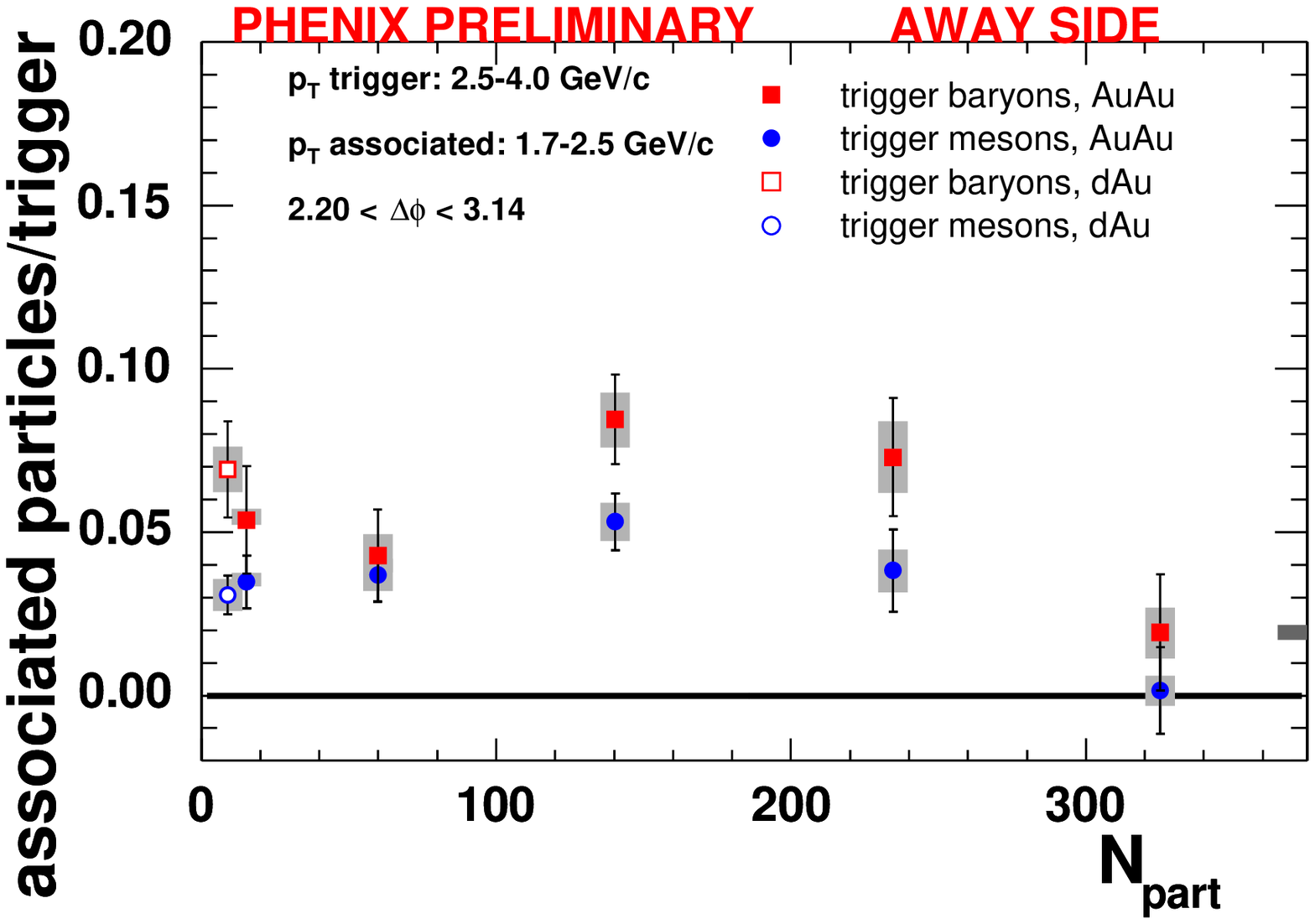}
\caption{Left: The near side associated yields as a function of centrality when the trigger particle
is identified as a meson or baryon. See the text for a discussion of the theory curves. Right: The away 
side associated yields for meson and baryon trigger particles.}
\label{fig:identified_trig}
\end{figure}

To investigate the production mechanism for the excess baryons observed at 2-5 GeV/c in central Au+Au 
collisions, PHENIX has studied azimuthal two-particle correlations with one of the particles identified 
as a baryon or meson~\cite{identified_trigger}. 
If the intermediate $p_T$ trigger particles are produced by fragmentation, they should show 
correlations with other particles in the jet. If they are produced by coalescence of flowing 
thermal partons, they should not show correlations with lower momentum particles. 
Figure~\ref{fig:identified_trig} shows that the near side associated yields per trigger are very similar
for meson and baryon triggers. The theory curves\cite{fries} show the expected behavior for protons 
(dashed) and pions (dot-dashed) for coalescing thermal quarks. Also shown is the away side yield per 
trigger particle,
which shows a larger yield on the away side when a baryon is the trigger particle.
The results seem to be inconsistent with models in which medium $p_T$ hadron production is dominated
by thermal quark coalescence. Models which allow coalescence of thermal partons with jet fragments 
may do better.

\begin{figure}
\center
\includegraphics[width=0.50\textwidth]{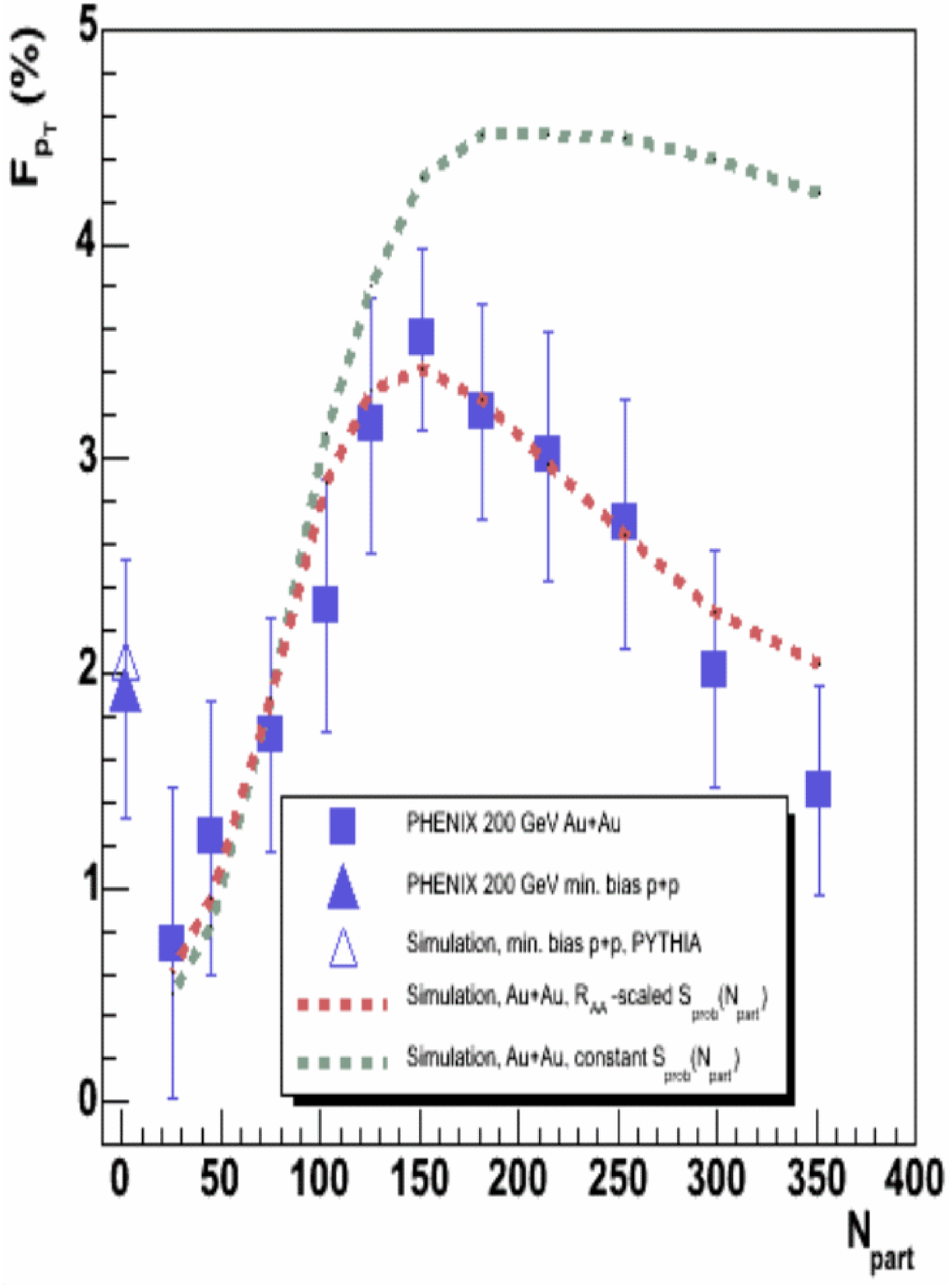}
\caption{Measured fluctuations in the average $p_T$ versus centrality for Au+Au collisions. $F_{PT}$ 
is the percent difference between the standard deviations of the average $p_T$ of the data and the 
average $p_T$ of a mixed event sample. See the text for a description of the curves.}
\label{fig:pt_fluct_plot}
\end{figure}

PHENIX has observed small nonrandom fluctuations in event-by-event averaged $p_T$ values from 200 GeV
Au+Au collisions. A PYTHIA based simulation study has shown that their $p_T$ and centrality dependence 
can be explained if they are caused by correlations due to jets, when proper account is taken of the 
measured jet suppression factor~\cite{pt_fluctuations}. Figure~\ref{fig:pt_fluct_plot} shows the measured
fluctuations from Au+Au data compared with a model made by embedding high $p_T$ PYTHIA events into 
simulated AuAu events. When the model takes account of the observed suppression of pions in central
Au+Au collisions (lower curve), it agrees quite well with the data.

HBT radii have been measured by PHENIX from pion-pion, kaon-kaon and proton-proton
correlations produced in 200 GeV Au+Au collisions. The fits to the correlations have been made with 
the so-called partial coulomb corrections (which allow for the fact that the data contain some
particles that were not directly produced in the collision). The results, which extend out to a $k_t$ of 
1.2 GeV, show that all three of the hadron pairs studied yield consistent radii, and the radii fall with 
$m_t$ as
expected for collective flow. The ratio of $R_{out}/R_{side}$ is consistent with one, implying a short
emission time that seems inconsistent with expectations if a QGP is formed~\cite{phenix_hbt}.

\begin{figure}
\center
\includegraphics[width=0.80\textwidth]{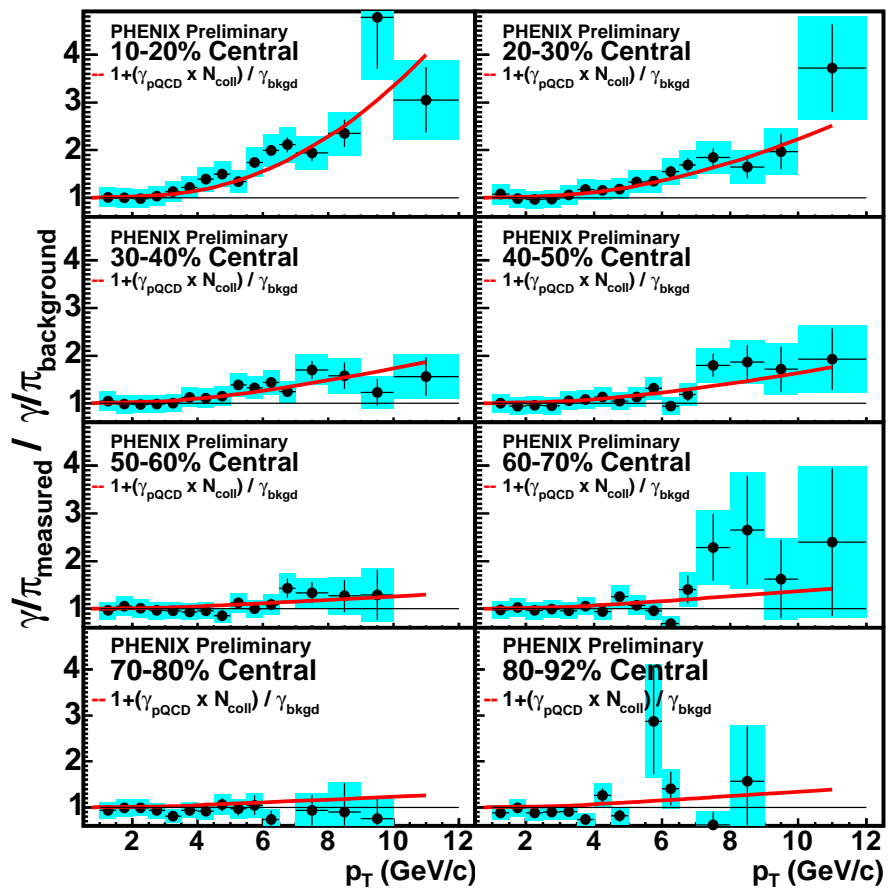}
\caption{Direct photon signal for 200 GeV Au+Au collisions as a function of centrality. See the text
for a description of the double ratio. The heavy solid curve is a pQCD estimate.}
\label{fig:direct_photons}
\end{figure}

Figure~\ref{fig:direct_photons} shows a summary of PHENIX direct photon yield measurements for 200 GeV 
Au+Au collisions~\cite{phenix_photons}, compared with a binary collision scaled pQCD estimate. 
These first direct photon measurements from 
RHIC are presented as a measured ratio of photon yield to $\pi^0$ yield, divided by the
ratio of the expected photonic meson decay yield (based on measurements) to the $\pi^0$ yield. 
The use of this double ratio causes some experimental systematic errors to cancel. 
Any excess over 1 observed in this ratio is the number of direct photons relative to the number of (background)
photons due to meson decays. The results show that direct 
photon production in 200 GeV collisions is well described by pQCD - ie. unlike hard partons,
hard photons are not suppressed in central Au+Au collisions at RHIC. These data, from Run 2, are
not yet precise enough to allow any conclusions about possible thermal direct photons at RHIC.

\begin{figure}
\center
\includegraphics[width=0.90\textwidth]{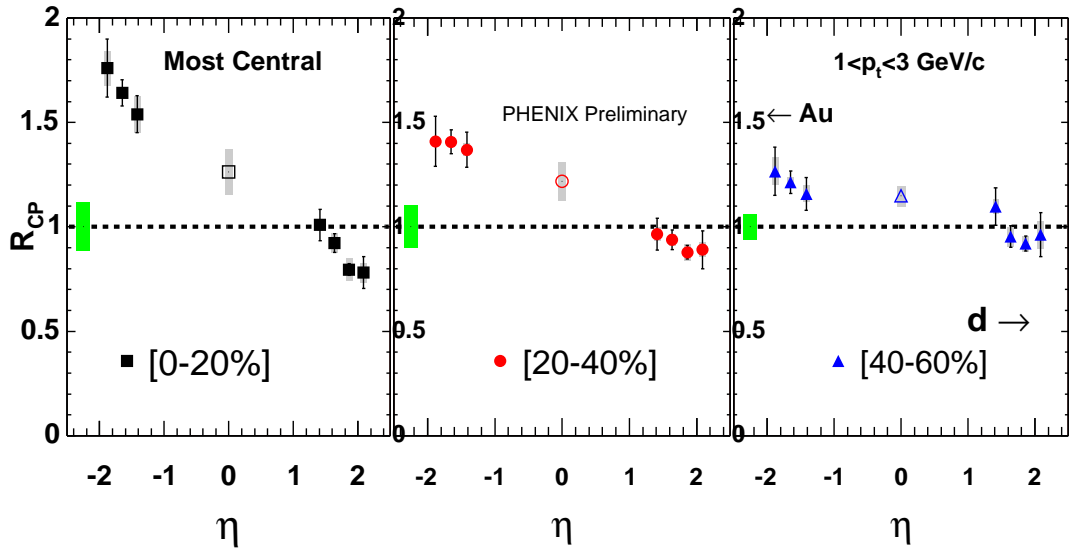}
\caption{The ratio of binary collision scaled central to peripheral yields for charged hadrons in 
200 GeV Au+Au, as a  
function of rapidity for three centrality bins. The yields inside $1<{\eta}<2$ were measured using 
punch-through hadrons in the muon identifiers.}
\label{fig:muon_punchthrough_rcp}
\end{figure}

PHENIX has recently developed new techniques for measuring hadrons in the muon arms. These techniques
have been used~\cite{phenix_muon_rcp} to study the nuclear modification factor for hadrons in 
d+Au collisions in the 
range $-2 < \eta < 2$. Figure~\ref{fig:muon_punchthrough_rcp} shows the ratio of central to peripheral 
yields ($R_{CP}$) for hadrons in three centrality bins in the central and muon arms. The observed
suppression on the deuteron-going side (positive rapidity) and enhancement on the Au-going side (negative
rapidity) are qualitatively consistent with parton shadowing (or saturation) at small x in the 
Au nucleus, and anti-shadowing at large x, all superimposed on a Cronin enhancement.

\begin{figure}[htb]
 \centering
  \includegraphics[width=0.60\textwidth]{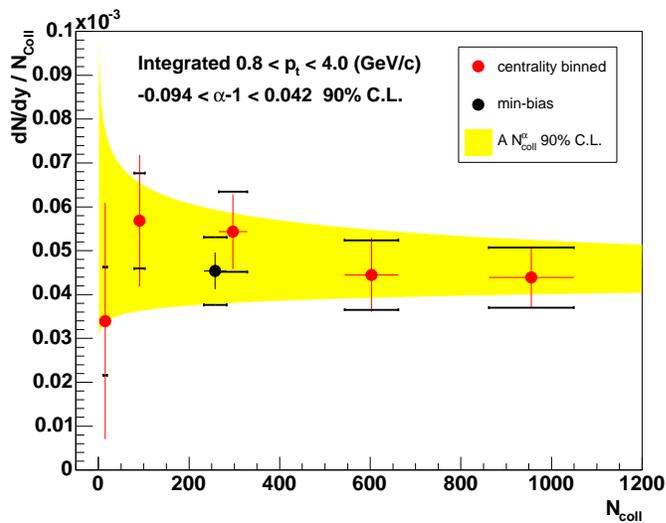}
  \caption{Open charm integrated yields per binary collision for 200 GeV Au+Au.}
  \label{fig:auau_dndy}
\end{figure}

The observation of strong suppression of high $p_T$ light quark hadrons at 200 GeV in central Au+Au 
collisions at RHIC has led to interest in measuring the spectral shape for heavy quark mesons.
PHENIX has measured single electron spectra at midrapidity in 200 GeV p+p, d+Au and Au+Au 
collisions~\cite{phenix_open_charm} as a means of studying open charm (D meson) and open beauty 
(B meson) production in these reactions. The electron spectra, after subtraction of electrons from
photon conversion and light quark decays, are assumed to be due to heavy quark decays. 
The d+Au electron spectra are consistent, within errors, with the binary scaled p+p 
spectra. This indicates that there is no strong initial state modification of the gluon distribution 
function in Au in the x range $10^{-2} - 10^{-1}$. For Au+Au the 
integrated dN/dy values are consistent with binary scaling within errors (see Figure~\ref{fig:auau_dndy}), 
but nothing can be concluded yet about the energy loss of heavy quarks in central Au+Au collisions, 
given the precision of the Au+Au data from Run 2. 

\begin{figure}[htb]
 \centering
  \includegraphics[width=0.60\textwidth]{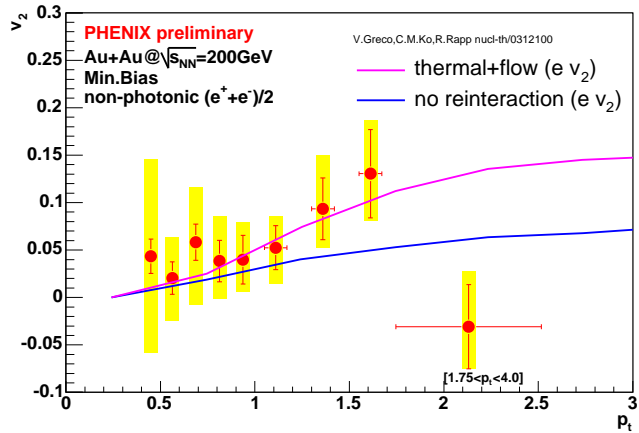}
  \caption{Open charm $v_2$ for 200 GeV Au+Au collisions, compared with model
estimates.}
  \label{fig:charm_v2}
\end{figure}

The $p_T$ spectrum alone at medium $p_T$ does not discriminate well between a scenario 
where the charm quark has minimal interaction with the medium, and a scenario where the charm quark 
interacts strongly with the medium and participates in hydrodynamic flow. The measurement of 
the elliptic flow parameter ($v_2$) for
open charm offers the hope of discriminating between these two scenarios~\cite{phenix_open_charm}. 
PHENIX has studied event anisotropy using $\pi^0$, photon and electron data~\cite{charm_v2}.  
Figure~\ref{fig:charm_v2} shows the 
result of a preliminary analysis of Run 2 Au+Au data to extract the $v_2$ for non-photonic single 
electrons (predominantly due to open charm in the $p_T$ range covered by the data). The measurement is 
compared with a prediction assuming no interaction of charm with the medium, and a prediction 
assuming thermalization and flow of the charm. The results are inconclusive at the level of precision 
available from the Run 2 data, but the Run 4 data set is expected to provide two orders of magnitude 
higher electron yields, as well as single muon measurements.

\begin{figure}[htb]
 \centering
  \includegraphics[width=0.50\textwidth]{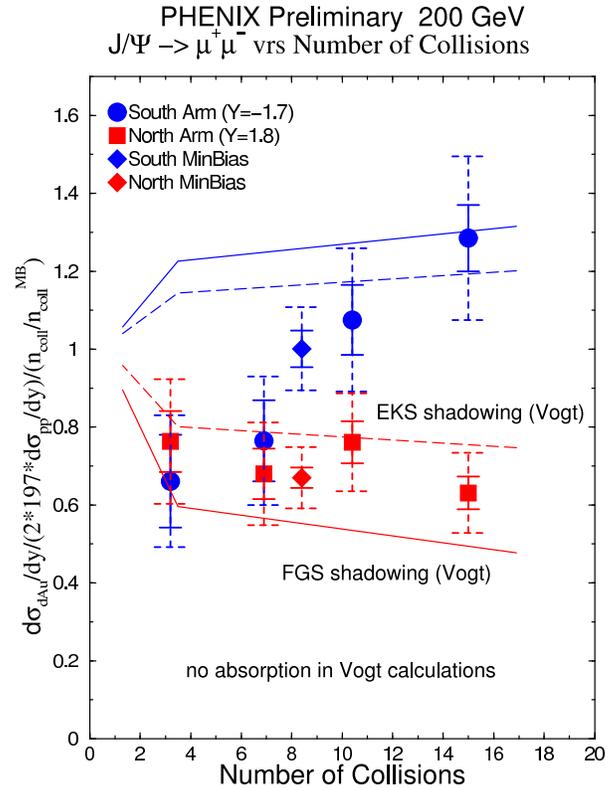}
  \caption{Centrality dependence of the nuclear modification factor for 200 GeV d+Au collisions. See
the text for discussion.}
  \label{fig:jpsi_rda}
\end{figure}

PHENIX has measured $J/\psi$ production at forward, middle and backward rapidities in 200 GeV p+p and 
d+Au collisions~\cite{phenix_jpsi}. The p+p total cross section is extracted, providing the baseline 
$J/\psi$ production rate in the absence of nuclear effects. The centrality dependence of the $J/\psi$ 
yield per binary collision in d+Au collisions contains information about how the cross section is 
modified by the cold nuclear matter present in d+Au collisions. Figure~\ref{fig:jpsi_rda} shows the 
measured nuclear modification factor ($\sigma_{d+Au}/(N_{coll}~\sigma_{p+p})$) in d+Au as a function 
of the number of binary collisions in the two muon arms, along with some model predictions of the 
effects of shadowing and anti-shadowing. At low x 
in the Au nucleus (forward rapidity), the centrality dependence is weak and qualitatively consistent 
with the model predictions. However at large x in the Au nucleus the nuclear modification factor 
rises strongly with increasing centrality. This is not yet understood.

PHENIX has made first measurements aimed at comparing the $\phi \rightarrow e^+e^-$ (120 events) and 
$\phi \rightarrow K^+K^-$ decays (207 events) for 200 GeV d+Au collisions~\cite{dau_phi}. Within the
fairly large errors, the yield and transverse mass distributions agree for the two channels.

Finally, there has been a search for the anti-pentaquark via the $\bar{\Theta} \rightarrow \bar{n}~K^-$
channel, where the $\bar{n}$ can be detected using its annihilation in the electromagnetic 
calorimeter~\cite{theta_search}. The search uses the 30\% most peripheral d+Au events. 
At the time of the conference a statistically significant mass peak was 
seen at an invariant mass of 1.54 GeV. But, after an independent analysis did not see the peak, it
was found that the original analysis lacked a necessary timing correction which, when applied, caused the
peak to fall below statistical significance.


\begin{thebibliography}{99}

\bibitem{phenix_highpt}
Klein-Boesing~C on behalf of the PHENIX Collaboration, these proceedings; nucl-ex/0403024.

\bibitem{phenix_pid}
Matathias~F on behalf of the PHENIX Collaboration, these proceedings; nucl-ex/0403029.

\bibitem{phi_rcp}
Kotchetkov~D on behalf of the PHENIX Collaboration, these proceedings.

\bibitem{jets}
Rak~J on behalf of the PHENIX Collaboration, these proceedings.

\bibitem{STAR_awaysidejet}
Adler~C~et al, \PRL {\bf 90} 082302

\bibitem{identified_trigger}
Sickles~A on behalf of the PHENIX Collaboration, these proceedings; nucl-ex/0403028.

\bibitem{fries} Fries~R~J, M\"{u}ller~B, Nonaka~C and Bass~S~A, 2003 {\sl Phys.\
Rev.\ C} {\bf 68} 044902

\bibitem{pt_fluctuations}
Tannenbaum~M~J on behalf of the PHENIX Collaboration, these proceedings; nucl-ex/0403048.

\bibitem{phenix_hbt}
Heffner~M on behalf of the PHENIX Collaboration, these proceedings.

\bibitem{phenix_photons}
Frantz~J on behalf of the PHENIX Collaboration, these proceedings.

\bibitem{phenix_muon_rcp}
Liu~M~X on behalf of the PHENIX Collaboration, these proceedings; nucl-ex/0403047.

\bibitem{phenix_open_charm}
Kelly~S on behalf of the PHENIX Collaboration, these proceedings; nucl-ex/0403057.

\bibitem{charm_v2}
Kaneta~M on behalf of the PHENIX Collaboration, these proceedings.

\bibitem{phenix_jpsi}
DeCassanac~R~G on behalf of the PHENIX Collaboration, these proceedings; nucl-ex/0403030.

\bibitem{dau_phi}
Seto~R on behalf of the PHENIX Collaboration, these proceedings; nucl-ex/0404002.

\bibitem{theta_search}
Pinkenburg~C on behalf of the PHENIX Collaboration, these proceedings; nucl-ex/0404001.


\end{thebibliography}
\end{document}